\documentclass{article}

\usepackage{amssymb,amsmath,amsfonts,amsthm}
\usepackage{graphicx}
\usepackage{bbm}

\usepackage{hyperref}
\usepackage{graphicx}
\usepackage{pst-all}
\usepackage{graphicx}
\usepackage{graphics}
\usepackage{epsfig}
\usepackage{multicol}
\usepackage{color}
\usepackage{epsf}

\usepackage{rotating}

\renewcommand{\c}[1]{\mathcal{#1}}

\newcommand{\<}{\langle}
\renewcommand{\>}{\rangle}

\begin{document}

\begin{center}
{\LARGE Composition of quantum operations \\
 and products of random matrices} \\[12pt]
Wojciech~Roga$^1$, Marek~Smaczy{\'n}ski$^1$ and Karol~{\.Z}yczkowski$^{1,2}$
\end{center}

\medskip
\noindent
$^1$Instytut Fizyki im.~Smoluchowskiego,
Uniwersytet Jagiello{\'n}ski,
PL-30-059 Krak{\'o}w, Poland \\
$^2$Centrum Fizyki Teoretycznej, Polska Akademia Nauk,
PL-02-668 Warszawa, Poland

\bigskip \bigskip
\hskip 4.0cm {\sl dedicated to the memory of Ryszard Zygad{\l}o}
\bigskip \bigskip

\noindent
Email: \texttt{<wojciech.roga@uj.edu.pl>}, \
\texttt{<log{\_}marco@poczta.onet.pl>}, \\
\texttt{<karol@tatry.if.uj.edu.pl>} 

\medskip\noindent
May 6, 2011

\bigskip
\noindent
\textbf{Abstract:}
Spectral properties of evolution operators corresponding to random maps and quantized
chaotic systems  strongly interacting with an environment can be described by the ensemble of 
non-hermitian random matrices from the real Ginibre ensemble. We analyze evolution operators
$\Psi=\Psi_s \cdot \dots \cdot \Psi_1$ representing the composition of $s$ random maps 
and demonstrate that their complex eigenvalues are 
asymptotically described by the law of Burda et al. obtained for a product
of $s$ independent  random complex Ginibre matrices.
Numerical data support the conjecture that the same results are
applicable to characterize the distribution of eigenvalues of 
the $s$--th power of a random Ginibre matrix.
Squared singular values of $\Psi$ are shown to be 
described by the Fuss-Catalan distribution of order $s$.
Results obtained for products of random Ginibre matrices
are also capable to describe the $s$--step evolution operator
for a model deterministic dynamical system 
-- a generalized quantum baker map subjected to
 strong interaction with an environment.

\medskip
\noindent

\section{Introduction}


Under the assumption of classical chaos
the corresponding unitary quantum evolution,
representing dynamics of an isolated quantum system,
can be described \cite{S99,H06}  by random unitary matrices
of the circular ensembles of random matrices \cite{Me04}.
If the quantum system is not isolated,
but it is coupled to an environment, its time evolution is not unitary. 
In such a case one describes the quantum state by a density operator $\rho$,
which is Hermitian, $\rho=\rho^{\dagger}$,
positive, $\rho \ge 0$, and normalized, Tr$\rho=1$.
The time evolution of such a system can be described by master equations \cite{AL87},
or by quantum operations ~\cite{BZ06},
 which correspond to a stroboscopic picture and discrete dynamics.

A quantum operation is described by a {\sl superoperator} $\Psi$, 
which acts on the space of density operators.
Let $N$ denotes the size of a density matrix $\rho$.
Then the superoperator is represented by a matrix $\Psi$ of size $N^2$. 
Such a matrix is in general not unitary,
but it obeys a quantum analogue of the Frobenius--Perron theorem \cite{BCSZ09},
so its spectrum is confined to the unit disk. 
Spectral properties of superoperators representing 
 interacting quantum systems were investigated
in~\cite{LPZ02,GMSS03,PCWE09}
and also analyzed in an NMR experiment ~\cite{WHEBSLC04}.
Similar properties exhibit also non-unitary evolution operators 
analyzed  earlier in context of quantum dissipative dynamics \cite{GHS88,Br01}.

Under the condition of classical chaos and strong decoherence 
the spectral properties of one-step evolution operators
of deterministic systems do coincide with these of random operations \cite{BCSZ09}
and can be described \cite{BSCSZ10} by the ensemble of non-hermitian Ginibre matrices.
All entries of such a random matrix are independent Gaussian variables.
Since a superoperator describing one-step evolution
operator can be represented as a real matrix~\cite{TDV00},
we are going to apply random matrices of
{\sl the real Ginibre ensemble}~\cite{LS91,SW08,AKxx}.

The main aim of this work is to study spectra of evolution operators
describing compositions of random quantum operations.
Furthermore, we analyze $s$--step evolution operators representing 
quantum systems periodically interacting with the environment
and compare statistical properties of complex spectra
with predictions of the theory of non-hermitian random matrices.

This work is organized as follows. In section 2
we recall necessary definitions of relevant ensembles 
of random matrices and briefly review recent results concerning
statistical properties of their products. Properties
of random maps and their compositions are analyzed in section 3.
The model deterministic dynamical system - a variant of the baker map
interacting with an environment is studied in section 4.

\section{Non-hermitian random matrices \\
            and their products}

Consider a random square matrix $G$ of size $N$
of the {\sl complex Ginibre ensemble} \cite{Gi65},
 generated according to the probability density
\begin{equation}
P(G)\propto\exp\left(-{\rm Tr} \; GG^{\dagger}\right).
\label{ginibre}
\end{equation}
This assumption implies that each entry $G_{mn}$
of the random matrix is an independent complex Gaussian variable 
of a fixed variance $\sigma^2=\xi^2/N$, where $\xi$ is a free parameter which sets the scale.
Eigenvectors of a random matrix $G$ from such an ensemble
are distributed according to the Haar measure on the unitary group,
while complex eigenvalues $z_i$ are described by the joint
probability distribution
\begin{equation}
P(z_1,\dots,z_N)\propto
\exp\left(-\sum_i|z_i|^2\right)  \prod_{i<j}|z_i-z_j|^2 \; .
\label{eigdistr}
\end{equation}
From this results one can evaluate the density of eigenvalues in the complex plane.
The density is rotationally symmetric and is a function of the moduls
$r=|z|$ of an eigenvalue \cite{Gi65,AKxx},
\begin{equation}
P(z) = \frac{1}{\pi} \frac{\Gamma(N, |z|^2)} {\Gamma(N)}
\label{densginue}
\end{equation}
where $\Gamma(s,x)$ denotes  the incomplete Gamma function,
\begin{equation*}
\Gamma(s,x)=\int_x^{\infty}t^{s-1}\exp(-t)dt.
\end{equation*}
In the asymptotic limit of large matrix size $N$
the level density becomes constant inside the disk of radius $R=\xi$,
and decays exponentially outside the disk. This fact, known as the circular law of Girko
\cite{Gi84}, is conveniently formulated under the normalization $\sigma^2=1/N$ so that $\xi=1$,
for which the spectrum of a random Ginibre matrix of a large dimension
covers uniformly the unit disk.

Several recent applications including  multiplicative diffusion processes \cite{GJJN03},
macroeconomic time series  \cite{BLMM07},
lattice gauge field theories \cite{LNW08} and
chiral ensembles of random matrices \cite{APS10,Ak11}
increased interest in statistical properties 
of products of non-hermitian random matrices \cite{CPV93}.
Let $Y$ denote a product of $s$ 
 independent square random matrices of size $N$ form the complex Ginibre ensemble,
 $Y=G_1G_2 \cdots G_s$.
 The density of the spectrum of $Y$ is rotationally invariant 
in the complex plane \cite{Burda10},
\begin{equation}
P(z)=\frac{1}{s\pi}\xi^{-2/s}|z|^{-2+(2/s)} \quad \text{for}\quad |z|\leq\xi.
\end{equation}
Here $\xi^2=\xi_1^2\xi^2_2...\xi^2_s$
denotes the product of scale parameters of each of the random matrices.
For simplicity we shall assume that all $s$ random matrices
are characterized by the same variance, so that $\xi=\xi_1^s$.
The radial density of the eigenvalues reads 
\begin{equation}
P(r)=\frac{2}{s}\xi^{-2/s} r^{-1+(2/s)} \quad \text{for}\quad r\leq\xi. 
\label{raddens}
\end{equation}
Based on the exact results for the level density for
the case $s=1$ of a single random matrix \cite{KS09,AKxx,KS10}
is was suggested by Burda et al \cite{Burda10_1}
to describe the finite size effects for the spectral density of
a product of $s$ random matrices by the dollowing ansatz, which
involves the complementary error function erfc$(x)=\frac{1}{\sqrt{2 \pi}}
\int_x^{\infty} \exp(-t^2/2)dt$

\begin{equation}
P_N(r)\equiv P(r)\; \frac{1}{2}{\rm erfc}\Big(q (r- \xi)\sqrt{N}\Big) .
\label{finite}
\end{equation}

Here $q$ is an adjustable parameter, which does not depend on the dimension $N$,
and the above form was reported \cite{Burda10_1,BJLNS11}
to describe well the data obtained numerically by diagonalization
of products of random Ginibre matrices.

\medskip 
Another way to describe a non-hermitian operator $A$ is to study its singular values. 
Their squares are equal to the eigenvalues of the 
positive matrix $AA^{\dagger}$.
To set the scale it is convenient to renormalize such a matrix  and define
\begin{equation}
W=\frac{AA^{\dagger}}{{\rm Tr} AA^{\dagger}}
\label{wishart}
\end{equation}
such that Tr$W=1$.

Let $\{\lambda_i \}$, $i=1,\dots, N$ denote the non-negative eigenvalues of $W$.
The normalization implies that their sum is equal to unity,
so if we use a rescaled variable,  $x_i=N \lambda_i$
its average value is equal to unity, $\langle x \rangle =1$.

If the matrix $A$ is random square Gaussian matrix from the Ginibre ensemble
the level density $P(x)$ describing the Wishart matrix $W$ given in (\ref{wishart})
is asymptotically (for a large matrix dimension $N$)
described by the Marchenko--Pastur distribution \cite{MP67}, 
\begin{equation}
FC_{1}(x) =\frac{1}{2\pi}\sqrt{\frac{4}{x}-1}\quad  \quad \text{for}
  \quad \quad 0\leq x\leq 4 .
\label{fc1}
\end{equation}
If $A$ is obtained as a product of $s$ independent square random Ginibre matrices,
$A=G_1G_2 \cdots G_s$, the level density of the Wishart-like matrix $W$
is given by the Fuss-Catalan distribution of order $s$ \cite{BBCC11,Ml10}.
The name of this distribution is related to the fact that
its moments are equal to the Fuss--Catalan numbers, 
often studied in combinatorics \cite{GKP}.
An explicit form of the FC distribution of order two, 
\begin{equation}
\!\!
FC_2(x) =\frac{\sqrt[3]{2} \sqrt{3}}{12 \pi} \;
 \frac{\bigl[\sqrt[3]{2} \left(27 + 3\sqrt{81-12x} \right)^{\frac{2}{3}} -
   6\sqrt[3]{x}\bigr] } {x^{\frac{2}{3}}
     \left(27 + 3\sqrt{81-12x} \right)^{\frac{1}{3}}},  
\label{fc2}
\end{equation}
valid for $ x \in [0,27/4]$, was derived first by Penson and Solomon \cite{PS01}
in context of construction of generalized coherent states.
More recently this formula was used in \cite{CNZ10,ZPNC10} 
to describe singular values of random quantum states,
the construction of which involves the product of two random matrices. 
Treating the sequence of Fuss-Catalan numbers as given
one can solve the Hausdorff moment problem
and find the corresponding probability distributions $FC_s(x)$.
They can be written down explicitly \cite{PZ11}
and  represented as a combination of the hypergeometric
functions  \cite{MOT} of the type ${}_sF_{s-1}$ of the same argument.
For instance, the Fuss--Catalan distribution of order three reads
\begin{eqnarray}
 FC_3(x) =\!\!\!\!\!\!&\frac{1}{\sqrt{2} \pi  x^{3/4}}\,
   _3F_2\left(-\frac{1}{12},\frac{1}{4},\frac{7}{12};\frac{1}{2},\frac{3}{4};\frac{27
   }{256}x\right)
-\frac{1}{4 \pi  x^{1/2} }\,
   _3F_2\left(\frac{1}{6},\frac{1}{2},\frac{5}{6};\frac{3}{4},\frac{5}{4};\frac{27
   }{256} x\right) \nonumber \\
&-\frac{1}{32 \sqrt{2} \pi  x^{1/4}}\,
   _3F_2\left(\frac{5}{12},\frac{3}{4},\frac{13}{12};\frac{5}{4},\frac{3}{2};\frac{27
   }{256}x\right).
\label{fc3}
\end{eqnarray}

The support of the
 Fuss--Catalan distribution $FC_s(x)$ of order $s$ is formed by an interval
$[0,(s+1)^{s+1}/s^s]$ \cite{BBCC11,CNZ10}.
Although this distribution was shown first to describe asymptotic
distribution of squared singular values of a product of $s$
independent Ginibre matrices, $A=G_1G_2 \cdots G_s$,
the same law describes asymptotically
the distribution of squared singular values of the $s$--th {\sl power}
of a square Ginibre matrix, $A=G^s$, \cite{AGT10}.

Analyzing discrete time evolution of Hermitian density matrices
in terms of quantum maps one copes with evolution operators
which can be represented by a real matrix \cite{BZ06}.
In particular, superoperators associated with random quantum maps \cite{BCSZ09}
can be described \cite{BSCSZ10} by real random matrices
of the Ginibre ensemble. These matrices can be formally defined
by the distribution (\ref{ginibre}) applied in the space 
of real matrices, so the argument of the exponent can be written as
Tr$AA^T$. To generate a random matrix pertaining to this ensemble
one takes $N^2$ independent random Gaussian variables of the same variance
and forms out of them a non-symmetric square matrix.

Statistical properties of the real Ginibre ensemble 
are more complicated to analyze \cite{LS91,SW08} than in the complex case.
 For instance, for the real Ginibre ensemble
the joint probability distribution of eigenvalues
depends explicitly on their imaginary parts, so the level density
is not rotationally invariant. In fact, there exist an
accumulation of eigenvalues along the real axis, which 
is compensated by the repulsion of complex eigenvalues 
in vicinity of the real axis \cite{LS91,AKxx}.
However, the number of real eigenvalues of a real random Ginibre 
matrix of size $N$ scales as $\sqrt{N}$, 
so that the non-uniform features of the spectrum can be neglected 
in the asymptotic  limit $N\to \infty$.
Also the distribution of singular values of products of
$s$ real Ginibre matrices can be described \cite{ZPNC10}
by Fuss--Catalan distribution of order $s$, 
originally applied \cite{BBCC11}
for products of complex Ginibre matrices.

\section{Random operations and their compositions}

A quantum operation is a linear map, $\rho\rightarrow\rho'=\Psi(\rho)$,
which sends the set of the density matrices
into itself, so it preserves positivity and trace of the input state $\rho$.
Any quantum operation acting on a $d$--dimensional state 
can be described by a unitary evolution $U$ applied to an extended system
followed by the  partial trace over the environment $\cal E$, 
\begin{equation}
\Psi(\rho)={\rm Tr}_{\cal E}\Big(U(\rho\otimes|\nu\>\<\nu|)U^{\dagger}\Big).
\label{lindb}
\end{equation}
Here $|\nu\>\in \c H_M$ denotes the initially pure state of the environment,
which is assumed to be $M$ dimensional, so the unitary matrix $U$
of size $dM$ acts on the composite Hilbert space ${\cal H}_d \otimes {\cal H}_M$.

If we reshape a density matrix $\rho$  of size $N$ into a vector of length $d^2$,
a quantum operation can be represented by a $d^2\times d^2$  matrix 
called superoperator.
It is convenient to represent a density operator $\rho$ of size $N$
by its Bloch vector,
\begin{equation}
\rho=\sum_{i = 0}^{d^2-1} a_i \; \gamma^i \ .
\label{eq2010}
\end{equation}
Here $\gamma^i$ denotes the set of $d^2-1$  
generators of the group $\textsf{SU(}d\textsf{)}$,
  which satisfy relations
${\rm Tr}\left(\gamma^i\gamma^j\right)=\delta^{ij}$, while $\gamma^0=1/\sqrt{N}$. 
As any density matrix is hermitian, $\rho=\rho^{\dagger}$,
the components of the Bloch vector are real,
 $a_i\in\mathbb{R}$ for $i=0,\dots,d^2-1$.
Thus the action of a quantum operation can be described
as an affine transformation on the Bloch vector $\vec{a}$ representing the
quantum state, 
$\vec{a}'=C \vec{a} +\vec{\kappa}$, where $C$ is a non-hermitian distortion
matrix of order $d^2-1$, while $\kappa$ is a translation vector 
of length $d^2-1$.
Using the Bloch vector representation one writes
the superoperator $\Psi$ as a real matrix, 
\begin{equation}
\Psi=\begin{bmatrix}
 1&0\\
\vec{\kappa}&C
\end{bmatrix} .
\label{struc}
\end{equation}
The spectrum of the non-symmetric matrix $C$ 
belongs to the complex plane.
Since the quantum operation $\Psi$ preserves the trace of the
density matrix, Tr$\Psi(\rho)={\rm Tr}(\rho)=1$,
the spectrum of the superoperator belongs to the unit disk.

Assuming that the matrix $U\in U(dM)$ in (\ref{lindb})
is taken randomly with respect  to the Haar measure
one obtains a random quantum operation \cite{BCSZ09}.
In such a case the spectrum of the associated
one-step evolution operator $\Psi$ was shown 
to consist of a single eigenvalue equal to unity,
corresponding to the unique invariant state, 
and the remaining part localized in the disk of radius $R=1/\sqrt{M}$
centered at the origin of the complex plane \cite{BSCSZ10}.
This characterization becomes exact for a large system size $d$.
Since the matrix $C$ is real, the spectrum  of $\Psi$ is symmetric
with respect to the real axis. As in the case of the real
Ginibre ensemble \cite{LS91,SW08,AKxx}
there exists a clustering of eigenvalues along the real axis,
but this effect vanishes in the limit $d\to \infty$.

The number $M$, which determines the size of the disk of complex eigenvalues, 
equal to the dimension of the auxiliary subsystem,
can be thus considered as a control parameter of the model describing the interaction of
the principal system with the environment.
Technically, $M$ determines the rank of the Hermitian dynamical matrix \cite{BZ06},
which describes the quantum operation.

\begin{figure}[h!tf]
\centering
\scalebox{.65}{\includegraphics{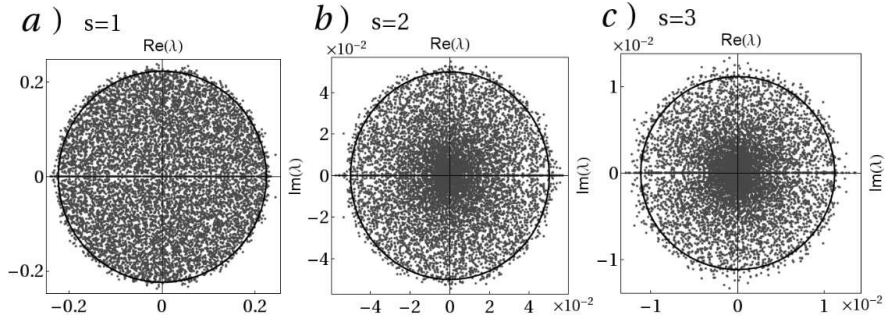}}
\caption{Superimposed spectra of 25 superoperators $C$ of dimension $d^2-1$
associated with random maps acting on density operators of dimension $d=20$ 
and obtained by an interaction with an environment of size $M=20$.
The superoperators represent a) single random maps, $s=1$; 
the composition of b) $s=2$ and c) $s=3$ random maps.
The disk of radius $R_s=1/\sqrt{M^s}$ (note the rescaling of both axes)
denotes the support predicted 
for the ensemble of random Ginibre matrices.
}
\label{fig1}
\end{figure}


\subsection{Spectral density of a random superoperator}

We analyzed spectra of evolution operators 
corresponding to compositions of random maps.
Let  $\Psi=\Psi_s \cdots \Psi_2 \cdot \Psi_1$,
where all $s$ random maps $\Psi_j$ are assumed to be
independent.  Figure \ref{fig1} shows the 
spectra of such operators for $s=1,2$ and $s=3$
for maps acting on a quantum system of size $d=20$.
To show the structure of the spectrum we magnified the
scale accordingly, letting the leading eigenvalue $z=1$
to remain outside the plot.
According to the prediction of the Ginibre ensemble
for $s=1$ the distribution of the spectrum is close to be uniform
in the disk of radius $R=1/\sqrt{M}$ (apart of the clustering of eigenvalues along the real axis),
while its structure changes for larger $s$.

Consider first the simplest case, $s=2$, in which two 
random operations (\ref{lindb}) act successively.
Assume that the first random operation $\Psi_1$
is due to an interaction with the environment ${\cal E}_1$ of dimension $M_1$, 
while the second operation $\Psi_2$
describes the interaction with the 
environment ${\cal E}_2$ of dimension $M_2$.
The resulting dynamics takes place in a tri--partite system
described in the Hilbert space ${\cal H}={\cal H}_p \otimes {\cal H}_1 \otimes {\cal H}_2$.
The first label $p$ refers to the principal system of dimension $d$,
while the other subsystems are labeled by the number of the environment ${\cal E}_1$
and ${\cal E}_2$.

After both operations the output state of the system, 
$\rho''=\Psi_2(\rho')=\Psi_2[\Psi_1(\rho)]$
can be obtained by a three-subsystem unitary evolution,
\begin{equation}
U=V_2V_1=(U_{p2}\otimes_1 {\mathbbm 1}_1)(U_{p1}\otimes_2 {\mathbbm 1}_2) ,
\label{U3}
\end{equation}
followed by the partial trace over the collective environment ${\cal E}_{12}$
of dimension $M=M_1M_2$. Note that the resulting unitary $U$ does not have a product structure, 
as the symbols  $\otimes$ represent the tensor product with respect to
two the different splitting of  the Hilbert space,
 ${\cal H}={\cal H}_{p1} \otimes_2 {\cal H}_2={\cal H}_{p2} \otimes_1 {\cal H}_1$. 

\begin{figure}[h!tf]
\centering
\scalebox{.65}{\includegraphics{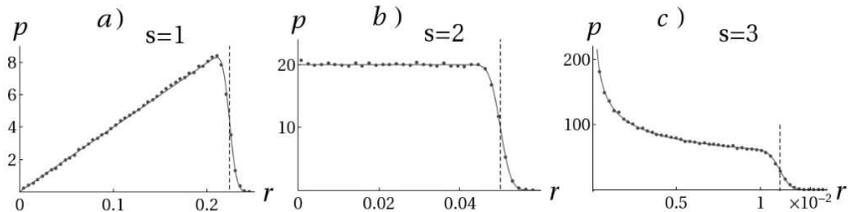}}
\caption{Radial density of complex eigenvalues of superoperators
 associated with a) a single random map, $s=1$
and a composition of b) $s=2$ and c) $s=3$ random maps.
Numerical data were obtained from a sample of 1000 superoperators 
 of dimension $d^2-1$ with $d=20$. 
The dimensions of all auxiliary subspaces are equal, $M=20$. 
Solid lines represent predictions (\ref{raddens})
for radial density for products of $s$  Ginibre matrices
with the correction (\ref{finite})  due to finite size effects.
Best fit gives the following values of the fitting parameter, 
a) $q\approx6$ for $s=1$ , b) $q\approx7$ for $s=2$, c) $q\approx68$ for $s=3$. 
Dashed vertical line represent the radius $R_s$ of the disk,
which determines the support of the essential spectrum 
in the limit $d\to \infty$.
}
\label{fig2}
\end{figure}

It is natural to expect that in the case of two random operations,
 $\Psi= \Psi_2 \cdot \Psi_1$,
the resulting effect will be similar as the one caused by a single interaction
with the combined environment  ${\cal E}_{12}$ of dimension $M=M_1M_2$.
This statement is equivalent to an assumption that the effect
of the action of the resulting unitary $U$ of the structure (\ref{U3})
is statistically indistinguishable
from the effect due to a global random unitary matrix $U_{p12}\in U(dM_1M_2)$
followed by the partial trace over  ${\cal E}_{12}$.
Our numerical results confirm that this approximation works fine,
as it implies that the subleading eigenvalues of the superoperator $\Psi$
live in the disk of radius $R_2=1/\sqrt{M}=1/\sqrt{M_1M_2}$.
   
The same reasoning shows that in the case of a composition of $s$
random operations, in which all dimensions of the environments are equal, 
 $M_1=M_2 =\dots=M_s=M$, 
the essential spectrum of the resulting superoperator is 
asymptotically confined  in the disk of  radius $R_s=R_1^s=1/\sqrt{M^s}$.
These predictions describe well the spectra of the 
evolution operators associated with the compositions
of two and three random maps and presented
in Fig. \ref{fig1} in panels $b$ and $c$, respectively.

The radial distribution of complex eigenvalues
collected of 1000 superimposed spectra
are presented in densities in Fig. \ref{fig2}. 
Numerical results can be described by the law (\ref{raddens})
obtained in \cite{Burda10} 
for a product of $s$ independent random complex Ginibre matrices.
To take into account the finite size effects
we used the ansatz (\ref{finite}) and  fitted the parameter $q$.

\subsection{Singular values of random superoperator}

\begin{figure}[h!tf]
\centering
\scalebox{.65}{\includegraphics{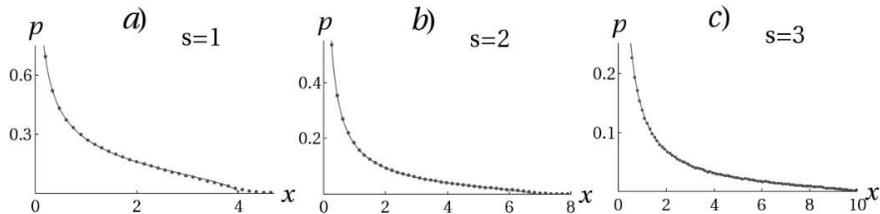}}
\caption{Density of normalized squared singular values, $x_i=(d^2-1)\lambda_i$,
of superoperators corresponding to a) a single random map, $s=1$;
a composition of b) $s=2$ and c) $s=3$ random maps
obtained for $d=20$ and $M=20$.
Numerical data collected from a sample of $1000$ random maps
are compared with the corresponding Fuss--Catalan distributions 
 (\ref{fc1},\ref{fc2}, \ref{fc3})
represented by solid curves.
}
\label{fig3}
\end{figure}

The distribution of squared singular values of a superoperator $\Psi$
associated with a single random map or the composition of $s$ of them
was analyzed numerically. The data obtained presented in Fig. \ref{fig3}
show a fair agreement with the Fuss-Catalan distribution of order $s$,
which describes properties of a product of 
$s$ complex Ginibre matrices \cite{BBCC11}.

\subsection{Average entropies}

To characterize the eigenvalue distribution $P(\lambda)$
of the positive matrix $\Psi\Psi^{\dagger}/ {\rm Tr} \Psi\Psi^{\dagger}$
one studies the Shannon entropy of the spectrum, 
$S=-\sum_{i=1}^{d^2-1} \lambda_i \ln \lambda_i$.
We put aside the leading eigenvalue $\lambda_1=1$
and renormalize the remaining $d^2-1$ eigenvalues so that their sum is set to unity.
The mean entropy computed numerically for a sample of $1000$ superoperators
$\Psi$ representing random maps with parameters $d=M=20$
reads 
$\<S\>_{\Psi}\approx-0.505$.
This value agrees with the asymptotic prediction for the Wishart matrices, 
$\<S\>_1=-\frac{1}{2}$ implied by the Marchenko-Pastur distribution (\ref{fc1}).
A similar agreement is obtained in the case $s=2$, for which the
 average entropy implied by the Fuss--Catalan distribution of order two (\ref{fc2})
 $\<S\>_2=-\frac{5}{6}\approx-0.833$ \cite{CNZ10},
while the numerical data give the average entropy $\<S\>_{\Psi}\approx-0.841$.
Numerical results for $s=3$ provide the value $\<S\>_{\Psi}\approx-1.093$,
whereas the Fuss-Catalan distribution $FC_3(x)$
leads to $\<S\>_3=-\frac{13}{12}\approx-1.083$.

\section{Generalized quantum baker map \\
    and $s$--step evolution operators}

To investigate statistical properties of
evolution operators associated with 
deterministic quantum systems interacting with an environment
we shall concentrate on a model dynamical system, 
the classical analogues of which is known to be chaotic.
Following the work of Balazs and Voros~\cite{BV89}
we consider the unitary operator describing 
the one--step evolution of the quantum baker map, 
\begin{equation}
B=F_d^{\dagger}\left[\begin{array}{cc}F_{d/2}&0\\0&F_{d/2}\end{array}\right] .
\label{baker}
\end{equation}
Here $F_d$ denotes the Fourier matrix of size $d$,
namely 
$[F_{d}]_{jk}=\exp({\sf i} j k / 2\pi d)/\sqrt{d}$ 
and it is assumed that the dimension $d$ of the Hilbert space ${\cal H}_d$ 
is even.

The standard quantum baker map $B$ may be generalized,
if in the definition of the unitary operator (\ref{baker})
the Fourier matrix $F_d$ is replaced by a two--parameter matrix 
$F_d^{\phi_1,\phi_2}$,
\begin{equation}
[F_d^{\phi_1,\phi_2}]_{jk}=\frac{1}{\sqrt{d}}
    \exp\Bigr[ {\sf i} \frac{  (j+\phi_1)( k+\phi_2)}{ 2\pi d} \Bigl]  ,
\label{Fphi}
\end{equation}
see Appendix D in \cite{POZ99}.
The choice of both phases in $[0,2\pi)$ 
does not influence the classical limit,
equal to the classical baker map.
Thus combining Eq. (\ref{Fphi}) and (\ref{baker})
one obtains a two-parameter
family of unitary quantum model dynamical systems,
which we denote by $B_{\phi_1,\phi_2}$.

A certain variant of a non-unitary baker map 
introduced by Saraceno and Vallejos is capable to
describe a dissipative quantum system \cite{SV96}.
Here we are going to investigate yet another model 
of non-unitary quantum baker map introduced in \cite{LPZ02,ALPZ04},
which is deterministic, conserves the probability,
and is capable to describe projective measurements
or a coupling with an external subsystem.
Such a nonunitary dynamics can be represented 
as a quantum map and written in its Kraus form \cite{BZ06},
\begin{equation}
\rho'= \Phi(\rho) = \sum_{j=1}^M  X_j \rho X_j^{\dagger} .
\label{Kraus}
\end{equation}
For any trace preserving operation the set of $M$ Kraus 
operators satisfies the identity resolution,
$\sum_{j=1}^M X_j^{\dagger} X_j ={\mathbbm 1}$.

The parameter $M$ is equal to the size of the environment
coupled to the principal system of the baker map.
Alternatively, $M$ can be interpreted as the 
number of different outcomes of a measurement process.
It is a free parameter of the model, which describes the
degree of the decoherence in the system introduced by the 
nonunitary map (\ref{Kraus}).

The model of the classical baker map is chaotic
and its dynamics can be
characterized by the dynamical entropy of Kolmogorov and Sinai
$H_{KS}$ equal to $\ln 2$~\cite{O02}.
To increase the degree of chaos one can simply take $L$
iterations of the classical system, for which the dynamical entropy reads
$L \ln 2$. This corresponds, in the quantum model,
to taking the $L$-th power of the unitary evolution $B$ given by 
(\ref{baker}), or its generalized version $B_{\phi_1,\phi_2}$
which involves  (\ref{Fphi}).
Increasing the parameter $L$ one increases the degree of chaos
in the classical model, and thus obtains unitary quantum operators
which the properties of which are well described \cite{SZ98}
by the Haar random matrices of the circular unitary ensemble.

\begin{figure}[htbp]
\centering
\includegraphics[width=1\textwidth]{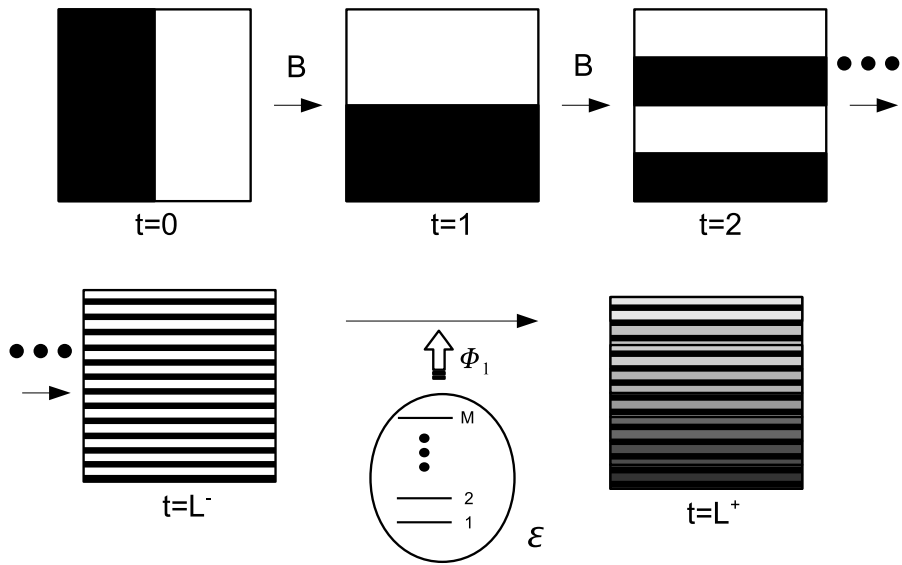}
\caption{ Sketch of the deterministic dynamical system - 
the generalized quantum baker map $\Phi_{M,L}$ --
analyzed in this work:
$L$ steps of the unitary dynamics 
followed by 
an interaction with an $M$-dimensional 
environment $\mathcal{E}$ described by the quantum operation $\Phi_1$.}
\label{fig_baker}
\end{figure}

In our model system we take the $L$--step unitary dynamics
of the baker map, $B^L$, followed by the non-unitary interaction
with the environment of dimension $M$. In other words the non-unitary
map $\Phi$ given by  (\ref{Kraus}) acts only every $L$ steps
of the unitary evolution. The structure of the complete
evolution operator is presented schematically in Fig. \ref{fig_baker}.

Thus the stochastic quantum baker map describing 
the non-unitary evolution of the generalized quantum 
baker map  \cite{LPZ02,ALPZ04} reads 
\begin{equation}
\Phi_{M,L}(\rho) =  \sum_{j=1}^M 
P_j [B^L \rho (B^{\dagger})^L] P_j^{\dagger} .
\label{qbaker}
\end{equation}
It consists of $M$ Kraus operators $P_j$,
which act on the unitarily rotated state  $B^L\rho (B^{\dagger})^L$.
It is assumed that the ratio $K=d/M$ is integer,
so one can decompose the Hilbert space ${\cal H}_d$ 
into the direct sum of $M$ mutually orthogonal
subspaces ${\cal H}_{(j)}$, $j=1,\dots, M$, 
of dimension $K$ each. Then the Kraus operator $X_j=P_jB^L$ 
is a projection operator onto the $K$--dimensional subspace 
${\cal H}_{(j)}$, so the sum 
$\sum_{j=1}^M X_j^{\dagger}X_j =\sum_{j=1}^M X_j$
is equal to identity, as required.
Thus the parameter $M$ in the non-unitary quantum baker map
studied in this section has a similar meaning than
$M$ parameterizing the random maps:
it describes the  degree of the interaction of the 
principal system with the environment.
Additionally, for each choice of the system parameters $(L,M)$
we may choose an  pair of phases $(\phi_1,\phi_2)$
which enter (\ref{Fphi}) and define the quantum model.
To obtain a better statistics we shall
superimpose spectra of superoperators 
obtained for fixed values of $(L,M)$
and various phases  $(\phi_1,\phi_2)$.

\begin{figure}[htbp]
\centering
 \includegraphics[width=1\textwidth]{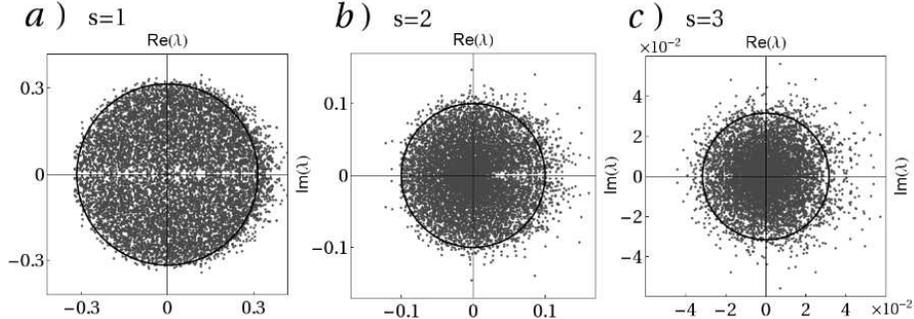}
\caption{Superimposed spectra of $60$  superoperators of the generalized quantum baker map 
acting on density operators of dimension $d=40$ and
characterized by parameters $L=20$ and $M=10$.
The superoperators represent a) single baker map (\ref{qbaker}), $s=1$; 
the s--step propagator $(\Phi_{M,L})^s$ for b) $s=2$ and c) $s=3$ time steps.
The disk of radius $R_s=1/\sqrt{M^s}$ (note the rescaling of both axes)
denotes the support predicted 
for the ensemble of random Ginibre matrices.}
\label{fig_bakera}
\end{figure}

Figure 5 presents exemplary spectra of superoperators $\Phi^s$
of the generalized quantum baker map
for $s=1,2,3$ obtained for fixed values of $L=20$ and $M=10$
and $60$ various pairs of phases $(\phi_1,\phi_2)$.
To display the structure of the bulk of the spectrum
the scale increases with the power $s$, 
so the leading eigenvalue $z_1=1$ is located outside the figure.
Apart of a few real eigenvalues, located for outside
the circle of radius $R_s=M^{-s/2}$,
the remaining eigenvalues are located close to the disk predicted
for products of random matrices in the asymptotic limit $d\to \infty$.
The spectra are symmetric with respect to the real axis
and exhibit the clustering of eigenvalues along the real
axis combined with the repulsion of eigenvalues in the vicinity
of the real axis. These effects, typical to the ensemble of real
Ginibre matrices \cite{LS91,SW08}, vanish in the asymptotic limit,
in which properties of products of complex and real random matrices
tend to coincide.

\begin{figure}[htbp]
\centering
 \includegraphics[width=1\textwidth]{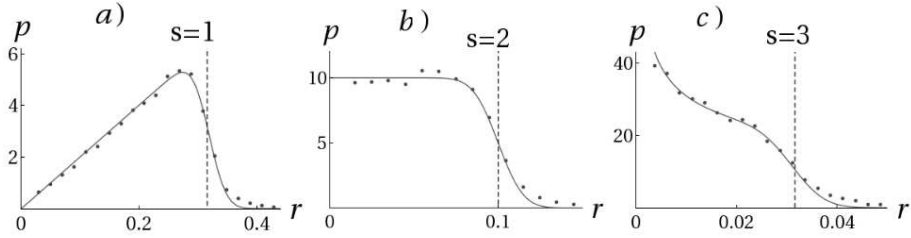}
\caption{Radial density of complex eigenvalues of superoperators
 associated with a) one step evolution operator of the baker map, $s=1$
and the $s$--step propagators for b) $s=2$ and c) $s=3$.
Numerical data were obtained from a sample of $100$ superoperators 
of the generalized quantum baker map 
acting on density operators of dimension $d=40$ and
characterized by parameters $L=20$ and $M=10$.
Solid lines represent predictions (\ref{raddens})
for radial density for products of $s$  Ginibre matrices
with the correction (\ref{finite})  due to finite size effects.
Best fit gives the following values of the fitting parameter, 
a) $q\approx 1.5$ for $s=1$, b) $q\approx 1.5$ for $s=2$, c) $q\approx 3.5$ for $s=3$. 
Dashed vertical line represent the radius $R_s$ of the disk,
which determines the support of the essential spectrum 
in the limit $d\to \infty$.}
\label{fig_bakerb}
\end{figure}

Figure 6 shows the radial density distribution $P(r)$
for complex eigenvalues of superoperators
of the generalized baker map $\Phi^s$ for $s=1,2,3$.
The dynamical parameters of the model are fixed, $d=40$, $L=20$ and $M=10$,
while to accumulate a necessary statistics we superimposed data
of $100$ superoperators obtained for different values  of the phases $(\phi_1,\phi_2)$.
Note that already for  $s=2$ the spectral properties of the superoperator $\Phi$ 
can be described by the ensemble of random Ginibre matrices.

To analyze properties of  superoperators associated with
$s$--step propagator of the generalized baker map,
we analyzed also statistical properties of squared
singular values of $\Phi^s$ equal to
eigenvalues of a positive operator $\Phi^s (\Phi^{\dagger})^s$.
Using the Kraus decomposition of a superoperator \cite{BZ06}, 
$\Phi=\sum_{i=1}^M X_i \otimes {\bar X_i}$
we find that  
$\Phi\Phi^{\dagger}=
\sum_{i,j=1}^M X_iX_j^{\dagger} \otimes {\bar X_i} X_j^T$.
Since in our model each Kraus operator is a projector
rotated by the same unitary matrix, 
$X_i=P_iU$, the unitaries cancel out. 
The projections operators are mutually orthogonal, $P_i P_j=\delta_{ij}P_j$
so the above expression reduces to a single sum
$\Phi\Phi^{\dagger}=\sum_{j=1}^M P_j \otimes P_j$.
Any operator $P_i$ projects onto the subspace of dimension $K=d/M$
so the operator $\Phi\Phi^{\dagger}=P$, where $P$ is a projector
on a space of dimension $N'=M(d/M)^2=d^2/M$,  
so its spectrum consists of $N'$ eigenvalues 
equal to unity and remaining $d^2(1-1/M)$ 
eigenvalues equal to zero - see Fig 7a.

Consider now the case $s=2$, in which we analyze 
the spectrum of $\Phi^2 (\Phi^{\dagger})^2$.
This operator can be written as 
$\Phi P \Phi^{\dagger}= (\Phi P) (\Phi P)^{\dagger}$.
Hence the singular values of $\Phi^2$ are
equal to the singular values of a truncated matrix 
$\Phi P$. Thus the generic, non-zero eigenvalues
of $W_2= \Phi^2 (\Phi^{\dagger})^2 / {\rm Tr}[ \Phi^2 (\Phi^{\dagger})^2]$
will be described by the Marchenko--Pastur distribution --
see Fig. 7b.
In a similar way, the singular values of $\Phi^s$
are equal to the singular values of a truncated matrix $\Phi^{s-1} P$
so its squared singular values
are described by the Fuss--Catalan distribution
of order $s'=s-1$.
Although the distributions ${\rm FC}_s(x)$
are known to describe the asymptotic distribution
of squared singular values of a product of complex Ginibre matrices,
they describe also statistical properties of $(s-1)$--step
propagators of the generalized baker map.

\begin{figure}[htbp]
\centering
 \includegraphics[width=1\textwidth]{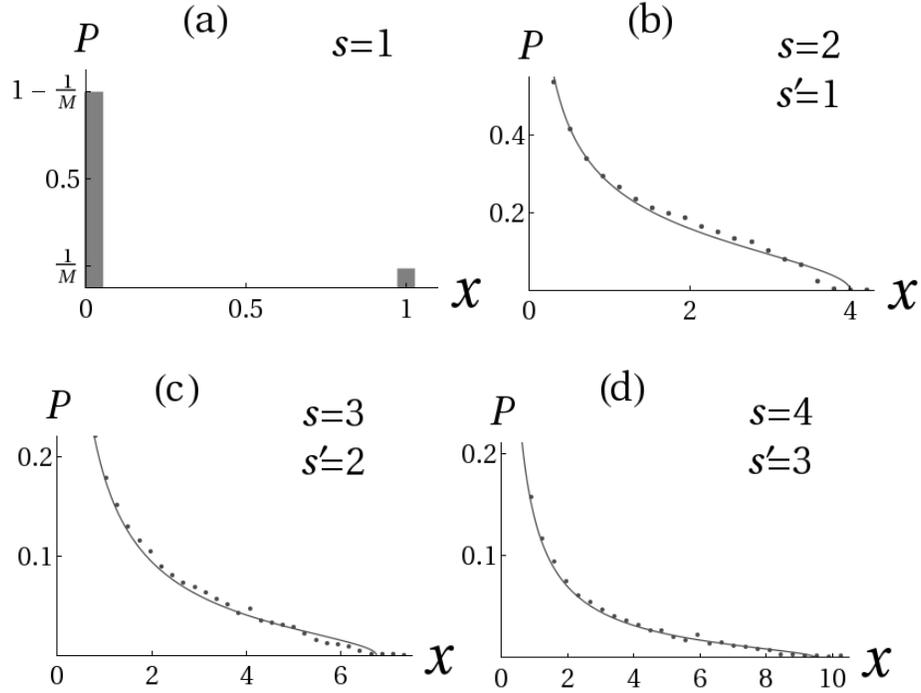}
\caption{Density of normalized positive squared singular values, $x_i=(d^2/M-1)\lambda_i$,
of superoperators corresponding to a) one step evolution operator
of the baker map, $s=1$;
the $s$--step propagators $(\Phi_{M,L})^s$ for b) $s=2$, c) $s=3$, 
and d) $s=4$ time steps (with the delta peak at $x=0$ removed).
Numerical data were obtained from a sample of $70$ superoperators 
of the generalized quantum baker map 
acting on density operators of dimension $d=40$ and
characterized by parameters $L=20$ and $M=10$.
Densities are compared with the corresponding Fuss--Catalan distributions 
of order $s'=s-1$
 (\ref{fc1}, \ref{fc2}, \ref{fc3})
  represented by solid curves. }
\label{fig_baker7}
\end{figure}

\section{Concluding remarks}

We analyzed complex spectra of superoperators  associated
to compositions of $s$ random maps.
Statistical properties of the eigenvalues can be
described by products of random matrices from the Ginibre ensemble.
Due to the fact that a quantum map $\Psi$ preserves hermicity of a
quantum state $\rho$, the superoperator $\Psi$ can be described \cite{BSCSZ10}
by an ensemble of {\sl real} Ginibre matrices,
for which a clustering of the eigenvalues along the real axis occurs  \cite{LS91}.
However, for large system sizes these finite--size effects can be neglected
and the density of eigenvalues can be compared with predictions
obtained for products of complex Ginibre matrices.
In particular, the radial density $P(r)$ of complex eigenvalues
of the superoperator associated with the composition $s$
random maps can be described by the algebraic law of Burda et al.
\cite{Burda10, Burda10_1},
while the distribution of the squared
singular values of the superoperator
(i.e. the eigenvalues of the positive
matrix $\Psi \Psi^{\dagger}$ are 
described by the Fuss--Catalan distributions of order $s$ \cite{BBCC11,CNZ10,ZPNC10,PZ11}.

Our numerical results support the conjecture that the
distribution of eigenvalues of a product of $s$
independent random Ginibre matrices, $G_1G_2 \cdots G_s$,
obtained in  \cite{Burda10_1,BJLNS11},
describe also the spectrum of $s$--th {\sl power} $G^s$
of a given random Ginibre matrix $G$.
This observation encouraged us to compare statistical properties
of $s$--step propagators of non-unitary quantum dynamical systems
with the predictions of random matrices.
Under the condition of strong classical chaos and  
sufficiently large coupling with the environment
the corresponding one--step
evolution operators can be described by the 
ensemble of real Ginibre matrices \cite{BSCSZ10}.

Investigating a generalized version of a model dynamical system --
the quantum baker map interacting with an environment --
we demonstrate that statistical properties of 
complex eigenvalues of $s$--step evolution operators associated
with such deterministic dynamical systems 
agree with predictions obtained for products of random Ginibre matrices.
For the dynamical system investigated the operator
$\Phi \Phi^{\dagger}$ is a projection operator 
with spectrum containing $\{0,1\}$, 
so it cannot be described by random matrices.
However, for a larger number $s$ of the time steps,
the squared singular values of $\Phi^s$
can be described by the Fuss--Catalan distribution 
of order $s'=s-1$
characteristic to the $s'$--th power of random matrices.

Thus products of random non-hermitian matrices, 
used to describe matrix valued diffusion  \cite{GJJN03} 
or random density operators \cite{CNZ10,ZPNC10},
can also be applied to characterize
statistical properties of multi-step evolution
operators corresponding to generic quantum 
dynamical systems strongly interacting with an
environment.

\bigskip 
Acknowledgements.
It is a pleasure to thank W. Bruzda, V. Cappellini, B. Collins,
I. Nechita, K. Penson and H.-J. Sommers for fruitful collaboration
and to Z. Burda and M.A. Nowak for an encouragement and
 helpful discussions. Financial support by the Transregio-12 project 
der Deutschen Forschungsgemeinschaft and
the special grant number DFG-SFB/38/2007 of
Polish Ministry of Science and Higher Education
is gratefully acknowledged.

\end{document}